\documentclass[nofootinbib,prd,preprintnumbers,superscriptaddress,aps]{revtex4}
\usepackage{amsmath}
\usepackage{amsfonts}
\usepackage{graphicx}
\usepackage[colorlinks=true,linkcolor=blue,citecolor=teal,urlcolor=blue]{hyperref}
\usepackage{xcolor}
\numberwithin{equation}{section}

\usepackage{hyperref}

\usepackage{physics}
\usepackage{bbold}
\usepackage{cases}
\usepackage{braket}
\usepackage[T1]{fontenc}
\usepackage{mathrsfs}
\usepackage{enumerate}
\usepackage{bm}
\usepackage{multirow}
\usepackage{comment}
\usepackage{mathrsfs}

\begin{document}

\title{Statistical physics on Euclidean Snyder space: connections with the GUP and cosmological implications}

\author{Boris Iveti\'c}
\email{bivetic@yahoo.com}
\affiliation{Hofferplatz 5/10, 1160 Vienna, Austria}
\author{Giuseppe Gaetano Luciano}
\email{giuseppegaetano.luciano@udl.cat}
\affiliation{Department of Chemistry, Physics and Environmental and Soil Sciences, Polytechnic School, University of Lleida, Av. Jaume II, 69, 25001 Lleida, Spain
}

\begin{abstract}
We develop a systematic formulation of statistical mechanics on Euclidean Snyder space, where noncommutativity is geometrically encoded in the curvature of momentum space. Adopting a realization-independent approach based on momentum-space invariants, we derive modified partition functions and thermodynamic quantities for systems obeying Maxwell-Boltzmann, Bose-Einstein and Fermi-Dirac statistics in both non-relativistic and ultrarelativistic regimes.
We show that momentum-space curvature induces temperature-dependent corrections that suppress the energy, entropy and energy density with respect to their standard counterparts. 
We apply these results to early-Universe cosmology, deriving the corresponding corrections to the Friedmann equations driven by the modified energy density of radiation. Using Big Bang Nucleosynthesis as a precision probe, we derive bounds on the Snyder deformation parameter and, via a phenomenological mapping, on the Generalized Uncerainty Principle (GUP) parameter, providing one of the most stringent cosmological and astrophysical constraints currently available.
Our analysis demonstrates that high-energy cosmological processes provide a predictive arena for testing momentum-space curvature and noncommutative geometry effects.
\end{abstract}

\date{\today}

\keywords {Noncommutative geometry, Snyder space, Statistical Thermodynamics, Cosmology}
\maketitle

\section{Introduction}
Noncommutativity provides a framework for generalizing the structure of spacetime at the smallest scales, playing a significant role in the investigation of its fundamental properties \cite{ref1, ref2, ref3, ref4, ref5}, and, more broadly, within quantum gravity \cite{Rovelli:1994ge,Maldacena:1997re, Kempf:1994su, Addazi:2021xuf, Hossenfelder:2012jw, Bosso:2023aht}. 
One of the earliest proposals along these lines is the Snyder model, originally introduced with the aim of formulating a version of quantum electrodynamics free from divergences \cite{ref6}. 
Since then, this construction has been explored in a wide range of physical settings, including classical mechanics \cite{ref7, ls2}, nonrelativistic quantum mechanics \cite{ref7, ref8, ref9}, classical electrodynamics \cite{ref10}, as well as scalar \cite{ref11, ref12, ref13} and gauge \cite{ref14, ref14bis, ref14ter, ref15, ref16, ref17, ref18} field theories. 
A characteristic feature of the Snyder construction is the presence of a noncommutativity  parameter, $\mathcal{P}$, interpreted as the radius of curvature of momentum space, whose inverse naturally defines a fundamental length scale in configuration space.

Despite its extensive applications, the Snyder construction has - up to a few notable exceptions \cite{ref19, ref20, iran3, iran4, Pachol2023, Pachwoj} - been only marginally investigated within the framework of statistical mechanics. This gap is not merely of formal interest, but may also entail potentially relevant experimental implications. Indeed, the energy scale $\mathcal{P}$ is expected to lie near the Planck mass, so that corrections induced by noncommutativity, typically scaling as $E/\mathcal{P}$, with $E$ denoting the characteristic energy of the process under consideration, are strongly suppressed. As a result, observable manifestations of noncommutative effects would require either extremely high energies or an exceptional level of experimental precision, thereby limiting the prospects for detection in terrestrial experiments in the foreseeable future. Interesting experimental proposals have been put forward, for instance, in \cite{Ter1, Ter2, Ter3}.

These considerations naturally motivate the study of Snyder-induced modifications in high-energy cosmological regimes, such as the early Universe. In particular, as we will show, the corrections arising from Snyder geometry exhibit a nontrivial temperature/energy dependence, indicating that the effects of noncommutativity could be significantly amplified during earlier cosmological epochs. It is worth noting that similar investigations have been carried out in alternative noncommutative frameworks, such as generalized uncertainty principle (GUP) models (see, e.g., \cite{cos1,cos2,cos3,cos4,cos5,cos6,cos7,Jizba:2009qf}).


%

Building on these premises, in this work we present a systematic statistical-mechanical analysis of Snyder space, deriving explicit temperature-dependent corrections to various thermodynamic quantities. We then explore the implications of these corrections in the context of Big Bang Nucleosynthesis (BBN), which allows us to place constraints on the deformation parameter of the Snyder algebra by requiring consistency with existing upper bounds on the fluctuations of the weak interaction freeze-out temperature. By connecting noncommutative geometry to a well-established observational framework, our study goes beyond purely formal considerations. It demonstrates that Snyder-induced modifications lead to potentially measurable effects in high-energy cosmological regimes, providing a concrete avenue to test and constrain models of quantum geometry. Overall, our results not only fill a notable gap in the literature but also highlight the predictive aspects of the Snyder framework.

The plan of the paper is as follows. In Sec. \ref{ESM}, we provide a brief introduction to the geometry of momentum space in the Euclidean Snyder model. In Sec. \ref{Stat}, we investigate the statistical mechanics and thermodynamics of the Snyder model for Maxwell–Boltzmann, Fermi–Dirac and Bose–Einstein distributions. In Sec. \ref{Cosm}, we derive the leading-order corrections to the Friedmann equations and discuss cosmological implications in the context of BBN. The paper concludes with a brief discussion and outlook in Sec. \ref{Conc}. Throughout this manuscript, we work in units where $\hbar=c=k_B=1$.

\section{The Euclidean Snyder model}
\label{ESM}

The Euclidean Snyder model is defined on a spherical momentum space, which can be most conveniently realized as a surface embedded in a four-dimensional Euclidean space:
\begin{equation}
\label{3surf}
\sum_{i=1}^{3} \pi_i^2 + \pi_4^2 = \mathcal{P}^2,
\end{equation}
where \(\pi_i\) are the three spatial momentum coordinates, \(\pi_4\) is the extra embedding coordinate in four-dimensional Euclidean space and \(\mathcal{P}\) denotes the maximal momentum scale of the Snyder model. 
Here, $\mathcal{P}$ denotes the radius of curvature of momentum space. To streamline the notation, from now on we will denote $\pi^2 \equiv \sum_{i=1}^{3} \pi_i^2$.

Before formulating a deformed dynamics, two ambiguities must be addressed. The first concerns the choice of three physical momentum coordinates, $p_i$, from the four coordinates of the embedding space, which, in principle, can be selected in infinitely many ways.
Any isotropic projection onto the $\pi_4 = 0$ plane can be considered:
\begin{equation}\label{pmus}
    p_i = g\!\left({\pi^2}/{\mathcal{P}^2}\right) \pi_i\,,
\end{equation}
with the inverse relation
\begin{equation}
    \pi_i = h\!\left({p^2}/{\mathcal{P}^2}\right) p_i\,,
\end{equation}

The function $g$ should be monotonic and satisfy $g(0)=1$, but is otherwise completely arbitrary. Different choices of $g$ in the literature are referred to as \emph{realizations}. The most commonly used are $g=\mathcal P (\mathcal{P}^2-\pi^2)^{-1/2}$ \cite{ref6} and $g=1$ \cite{Maggiore1993}. Different realizations give rise to distinct forms of the corresponding Heisenberg algebra. There is, however, no consensus on whether different realizations correspond to physically distinct models.

A second type of ambiguity arises naturally in any generalization of a standard theory. This concerns how the fundamental physical quantities, such as the mass or the Hamiltonian, as well as the basic physical equations, should be generalized. The primary constraint is provided by the correspondence principle: in the limit $\mathcal{P} \to \infty$, the standard theory must be recovered. Beyond this requirement, there remain infinitely many ways to define the deformed theory.

Remarkably, there exists a way to ensure that these two ambiguities exactly compensate each other. The only principle that needs to be postulated is that both physical quantities and physical equations should be expressed exclusively in terms of the \textit{geometric invariants} of momentum space. Following this prescription, let us consider, for example, the standard expression for the nonrelativistic kinetic energy:
\begin{equation}
E_k = \frac{p^2}{2m}.
\end{equation}
From a geometrical perspective, the term $p^2\equiv \sum_{i=1}^3p_i^2$ can be interpreted as the square of the geodesic distance between the point $(p_1,p_2,p_3)$ and the origin of momentum space. This provides the natural generalization of the kinetic energy for the spherical geometry of momentum space:
\begin{equation}\label{Ekin}
E_k = \frac{d^2(0,p)}{2m},
\end{equation}
where $d(p,0)$ denotes the geodesic distance between the point $p=(p_1,p_2,p_3)$ and the origin of momentum space.
Other physical quantities should be defined in an analogous manner. It follows that, by adopting this geometric principle, the resulting dynamics is independent of the specific choice of momentum coordinates \cite{ref9,ref13,Carmona2019}.

We therefore adopt the simplest possible “gauge” choice, $g=1$. Since the distributions considered in this work are isotropic in momentum space, we also employ a polar coordinate system:
\begin{equation}\label{intm}
d(0,p) = \mathcal{P}\, \omega, \quad \int d\Omega_p = 4\pi \mathcal{P}^3 \int_0^{\pi/2} \sin^2 \omega \, d\omega,
\end{equation}
where $\omega$ denotes the polar angle on the 3-sphere (\ref{3surf}), and the integration is restricted to $p_4 \ge 0$.
We stress that, for \(\mathcal P \to \infty\), the momentum space becomes  flat. Expanding at small \(\omega\) and setting \(p=\mathcal P\,\omega\), Eq.~(\ref{intm}) reproduces the standard Euclidean momentum-space integration.

By defining position operators in terms of the generators of the Lie algebra $so(4)$, i.e. $x_i=\frac{1}{\mathcal P} J_{i4}$, the deformed Heisenberg commutator of the Snyder model takes the form (see Appendix~\ref{AppA})
\begin{eqnarray}
\label{alg}
[x_i,x_j]&=&i\frac{J_{ij}}{\mathcal P^2}, \\[0mm] 
[x_i,p_j]&=&i\delta_{ij}\sqrt{1-\frac{p^2}{\mathcal P^2}}\,,
\label{alg2}
\end{eqnarray}
where \(J_{ij}=x_ip_j-x_jp_i\) is the angular momentum operator. As discussed in \cite{ref8}, this algebra introduces a fundamental scale set by $1/\mathcal{P}$. In the limit $\mathcal{P} \to \infty$, the deformed commutators reduce to the standard Heisenberg algebra, recovering ordinary quantum mechanics.


\section{Statistical mechanics on Snyder space}
\label{Stat}
In the context of statistical mechanics within the Snyder model, all the fundamental principles of conventional statistical mechanics remain fully applicable. Concepts such as a closed system, the criteria for its division into subsystems, the role of statistical distributions and related notions are entirely unchanged. Similarly, the definitions of thermal equilibrium, statistical independence and macroscopic quantities such as energy and entropy retain their standard meaning. Consequently, one can directly adopt the usual expressions from conventional statistical mechanics, with the sole modifications being the generalized definition of kinetic energy and the integration measure over momentum space, as given in (\ref{Ekin}) and (\ref{intm}) \cite{ref19, ref20, Pachol2023}.

Based on these considerations, we analyze how the curvature of momentum space affects the construction of partition functions, energy densities and other thermodynamic quantities. As a pedagogical starting point, we first consider a classical system obeying Maxwell–Boltzmann statistics in the non-relativistic regime. This choice allows us to introduce the key conceptual and computational modifications induced by a nontrivial momentum-space geometry in the simplest possible setting. The analysis is then extended to quantum systems obeying Bose–Einstein and Fermi–Dirac statistics, considering both the non-relativistic and ultrarelativistic regimes.

\subsection{Maxwell–Boltzmann distribution}
\label{SubMB}
Using the prescription \eqref{intm}, let us compute the generalized canonical partition function for a particle in Snyder space. By way of illustration, we focus on the case of a non-relativistic particle with energy $\epsilon(\omega) \approx m + \frac{\mathcal P^2 \omega^2}{2 m}$. Under this assumption, the partition function takes the form 
\begin{equation}
\label{MB}
    Z_{MB} = \frac{g V}{(2\pi)^3} 4\pi \mathcal{P}^3\int_0^{\pi/2} e^{-\frac{\mathcal{P}^2 \omega^2}{2 m T}} \sin^2 \omega \, d\omega,
\end{equation}
where \(V\) is the volume of configurational space, \(T\) is the temperature and the factor $g$ counts the internal degrees of freedom of the particle\footnote{We notice that the rest-mass contribution to the energy, being independent of the momentum variable $\omega$, gives rise to an overall multiplicative factor $e^{-m/T}$ in the partition function. Therefore, it can be consistently absorbed into a redefinition of the zero of energy.}. The above relation can be generalized to a system of many non-interacting particles by forming the $N$-particle partition function $Z^{(N)}_{MB}=Z_{MB}^N/N!$, which corresponds to the canonical ensemble with a fixed number of particles. Similarly, to account for a variable number of particles, one can consider the grand-canonical partition function
$ \mathcal{Z}_{MB} = \sum_{N=0}^{\infty} e^{\mu N/T} Z^{(N)}_{MB}= \exp\left( e^{\mu/T} Z_{MB} \right)
$, where the chemical potential $\mu$ controls the average number of particles in the system.

The integral \eqref{MB} can be evaluated as
\begin{equation}
Z_{MB}  = gV \sqrt{\frac{mT}{2\pi^3}} \frac{\mathcal P^2}{8}\, e^{-\frac{2mT}{\mathcal{P}^2}} \Bigg\{-2+ 2e^{\frac{2mT}{\mathcal P^2}}\operatorname{Erf} \left(\frac{\pi\mathcal P}{2\sqrt{2mT}}\right)+\operatorname{Erfc} \left(\frac{\pi \mathcal P^2-4imT}{2\mathcal P\sqrt{2mT}}\right) + \operatorname{Erfc} \left(\frac{\pi \mathcal P^2+4imT}{2\mathcal P\sqrt{2mT}}\right)\Bigg\}\,.
\end{equation}
The resulting expression, which involves error functions, is cumbersome and does not provide clear insight. 

To enable a meaningful comparison with the standard theory, we introduce the following definition:
\begin{equation}
\label{MBinf}
Z^{\infty}_{MB} = \frac{gV}{(2\pi )^3} 4\pi \mathcal P^3 \int_{0}^{\infty} e^{-\frac{\mathcal P^2\omega^2}{2mT}} \sin^2 \omega d\omega\,.
\end{equation}
As in standard statistical mechanics, extending the upper limit of integration to infinity introduces only exponentially suppressed corrections, due to the presence of the Gaussian factor. Therefore, this extension does not affect the result significantly. 
In the following, for simplicity of notation, we will omit the infinity symbol in $Z_{MB}^\infty$ while working with these extended integrals.

The integral \eqref{MBinf} reduces to the simpler expression
\begin{equation} \label{Z}
Z_{MB} = Z_{MB,0}\, \frac{\mathcal P^2}{2mT} \left( 1 - e^{-2mT/\mathcal P^2} \right),
\end{equation}
where $Z_{MB, 0}$ is the standard (flat momentum space) partition function,
\begin{equation}
Z_{MB,0}=gV \left(\frac{mT}{2\pi}\right)^{3/2}.
\end{equation}
Using the above relations, various thermodynamic quantities can be computed. In the physically relevant regime \(mT \ll \mathcal{P}^2\), the exponential term in \eqref{Z} can be expanded, and we retain only the leading-order contribution, which captures the dominant behavior of the system. For instance, in the canonical ensemble, the Helmholtz free energy is defined in the usual way in terms of the $N$-particle partition function as
\begin{equation}
F = - T \log Z_{MB}^{(N)} \approx -NT \log \left( \frac{eZ_{MB}}{N} \right),
\end{equation}
where, in the second step, we have applied Stirling’s approximation, $\log N!\approx N\log N-N$, which is valid for $N\gg1$. By using Eq. \eqref{Z}, we get
\begin{equation}
F = F_0 + \frac{mNT^2}{\mathcal P^2} + \mathcal O \left( \frac{m^2T^3}{\mathcal P^4} \right)
,
\end{equation}
where $F_0=-NT\log(eZ_{MB,0}/N)$ represents the standard free energy.  

In turn, the entropy takes the form
\begin{equation}
S = -\left(\frac{\partial F}{\partial T}\right)_{V,N} = S_0 - \frac{2mNT}{\mathcal P^2} + O \left( \frac{m^2T^2}{\mathcal{P}^4} \right),
\end{equation}
and the energy is given by
\begin{equation}
E = F + TS = E_0 - \frac{mNT^2}{\mathcal P^2} + \mathcal O \left( \frac{m^2T^3}{\mathcal P^4} \right),
\end{equation}
where, in both cases, the subscript zero denotes the corresponding expressions in standard statistical mechanics.
The last two expressions show that, within the Snyder model, 
both the entropy and the energy of a canonical ensemble following the Maxwell–Boltzmann distribution are lower than their corresponding standard values. This behavior can be understood as a consequence of the modified phase-space structure induced by the Snyder deformation, which effectively introduces a natural ultraviolet cutoff in momentum space. As a result, the number of accessible microstates is reduced compared to the standard case, leading to a suppression of entropy. A similar mechanism affects the energy, since high-momentum states, which would normally contribute significantly at high temperatures, are progressively less accessible in the deformed framework.

Finally, we consider the equation of state. From the definition of pressure, we obtain
\begin{equation}\label{pres}
P=-\left(\frac{\partial F}{\partial V}\right)_{T,N}=-\left(\frac{\partial F_0}{\partial V}\right)_{T,N}=\frac{NT}{V},
\end{equation}
which shows that the ideal gas equation of state does not depend on the geometry of momentum space. The same result was obtained in  \cite{iran4}

The above thermodynamic relations can be straightforwardly generalized to the grand-canonical ensemble by introducing the Landau (grand) potential $\Omega = -T \log \mathcal Z_{MB}$, which represents the natural thermodynamic potential for a system in thermal and chemical equilibrium with a reservoir. All relevant thermodynamic quantities, including the pressure, entropy, internal energy and the average particle number, can be consistently derived from $\Omega$ through the standard thermodynamic relations.


\subsection{Bose-Einstein and Fermi-Dirac distributions}

Following the approach developed in the previous section, the grand-canonical partition function for Fermi–Dirac and Bose–Einstein statistics is generalized to Snyder space in a straightforward manner as
\begin{equation}
\label{BEFD}
    \log\mathcal Z=\pm \frac{gV\mathcal P^3}{2\pi^2}\int_{0}^{\pi/2}
\sin^2\omega\;
\log\!\left[
1\pm e^{\frac{\mu-\epsilon(\omega)}{T}}
\right]
\, d\omega\,.
\end{equation}
As usual, the upper sign corresponds to fermions, while the lower sign corresponds to bosons. The above relation correctly reduces to Eq. \eqref{MB} in the classical limit $e^{-\left[\epsilon(\omega)-\mu\right]/T}\ll1$. 

From the grand-canonical partition function, the relevant thermodynamic quantities are obtained in the same way as in the standard case. For instance, the total average number of particles is defined as
\begin{equation}
\label{N}
    N=T\,\left(\frac{\partial \log \mathcal{Z}}{\partial \mu}\right)_{T,V}=
\frac{gV\mathcal P^3}{2\pi^2}
\int_{0}^{\pi/2}
\frac{\sin^2\omega\, d\omega}
{e^{\frac{\epsilon(\omega)-\mu}{T}}\pm 1}=\int dN_\omega\,.
\end{equation}
In turn, the total mean energy of the system is given by
\begin{equation}
\label{EBEFD}
E=\int_{N_{\omega=0}}^{N_{\omega=\pi/2}}\epsilon(\omega) dN_\omega=\frac{gV\mathcal P^3}{2\pi^2}
\int_{0}^{\pi/2}
\frac{\epsilon(\omega)\sin^2\omega\, d\omega}
{e^{\frac{\epsilon(\omega)-\mu}{T}}\pm 1}\,,
\end{equation}
while the definitions of entropy and pressure are
\begin{eqnarray}
    S&=& \frac{\partial }{\partial T}\left(T\log\mathcal Z   \right)_{V,\mu}=\frac{gV\mathcal P^3}{2\pi^2T}
\int_{0}^{\pi/2}
\frac{\left[\epsilon(\omega)-\mu\right]\sin^2\omega\, d\omega}
{e^{\frac{\epsilon(\omega)-\mu}{T}}\pm 1}+\log\mathcal Z\,,\\[2mm]
    P&=&T \left(\frac{\partial \log \mathcal Z}{\partial V}\right)_{T,\mu}=  \frac{T\log\mathcal Z}{V}=\pm \frac{gT\mathcal P^3}{2\pi^2}\int_{0}^{\pi/2}
\sin^2\omega\;
\log\!\left[
1\pm e^{\frac{\mu-\epsilon(\omega)}{T}}
\right]
\, d\omega\,.
\end{eqnarray}

Clearly, the explicit form of the single-particle energy $\epsilon(\omega)$ depends on the physical regime under consideration. Here, we focus on two particularly relevant limiting cases: 
\begin{itemize}
\item[\emph{i)}] in the \emph{non-relativistic regime}, the energy is given by $\epsilon(\omega) \approx m + \frac{\mathcal P^2 \omega^2}{2 m}$.
In this limit, the kinetic energy is small compared to the rest mass and thermal energies are typically much smaller than the particle mass (i.e., 
$m\gg T$). Under these conditions, all thermodynamic expressions reduce to the familiar Maxwell–Boltzmann formulas for classical particles, which can be directly derived from the results discussed in Sec. \ref{SubMB}.

\item[\emph{ii)}] On the other hand, in the \emph{ultrarelativistic regime}, the rest mass is negligible compared to the kinetic energy, and the single-particle energy simplifies to
$\epsilon(\omega) \approx \mathcal P \, \omega$. This regime is of particular relevance for the cosmological applications discussed below, as it characterizes the behavior of particles in the early Universe, where temperatures are extremely high and most species can be regarded as effectively massless. In this setting, it is natural to set the chemical potential to zero. Indeed, rapid number-changing processes - such as particle creation and annihilation occurring on timescales much shorter than the cosmic expansion - maintain chemical equilibrium, so that the thermodynamic properties of the system are fully determined by the temperature \cite{Landau,Kolbe}. 
\end{itemize}


\subsubsection{Ultrarelativistic regime}
Using Eq. \eqref{EBEFD}, the thermal energy density of an ultrarelativistic bosonic/fermionic gas in thermal equilibrium takes the form
\begin{equation}
\rho=\frac{g\mathcal P^4}{2\pi^2} \int_{0}^{\infty} \frac{\omega\sin^2 \omega \, d\omega}{e^{\mathcal P \omega /T} \pm 1},
\end{equation}
where, as discussed below Eq.~\eqref{MBinf}, extending the integral to infinity does not appreciably affect the result.
In the limit where the momentum scale $\mathcal P$ is much larger compared to the temperature $T$, the leading-order behavior is provided by
\begin{eqnarray}
\label{densB}
\rho_{BE}&=&\rho_{BE,0}\left[1-\frac{40\pi^2}{63}  \frac{T^2}{\mathcal{P}^2} +\mathcal{O}\left(\frac{T^4}{\mathcal P^4}\right) \right]\\[2mm]
\rho_{FD}&=&\rho_{FD,0}\left[1-\frac{310\pi^2}{441}  \frac{T^2}{\mathcal{P}^2} +\mathcal{O}\left(\frac{T^4}{\mathcal{P}^4}\right) \right],
\label{densF}
\end{eqnarray} 
for bosons and fermions, respectively. 
where   
\begin{equation}
\rho_{BE,0}=\frac{g\pi^2}{30}T^4\,, \qquad \rho_{FD,0}=\frac{7g\pi^2}{240}T^4\,,
\end{equation}
are the standard boson and fermion energy densities. The latter are consistently recovered in the limit $\mathcal P\rightarrow\infty$.

\section{Cosmological implications}
\label{Cosm}

In this section, we investigate the cosmological implications of a curved momentum space. In particular, we analyze how the modifications to the phase–space structure induced by momentum–space curvature affect the thermodynamic properties of relativistic matter and, consequently, propagate into the Friedmann equations. As shown in the previous section, these corrections alter the energy density of the cosmic fluid, leading to deviations from the standard cosmological dynamics at high energies. We focus on how such effects become relevant in the early Universe and discuss the resulting impact on the expansion history. 

Toward this end, we first briefly review the derivation of the modifications to the Friedmann equations induced by the emergence of a minimal length/area within the framework of the Generalized Uncertainty Principle (GUP). In this approach, quantum–gravity–motivated deformations of the canonical commutation relations lead to a modified phase–space structure, which in turn affects the thermodynamic properties of matter fields. When applied to a cosmological setting, these modifications translate into corrections to the energy density and pressure entering the Friedmann equations, resulting in deviations from the standard expansion dynamics at high energies.

This review serves a twofold purpose. On the one hand, it provides a well-established reference framework in which the impact of a minimal length/area quantization on cosmological evolution has already been extensively studied. On the other hand, it prepares the ground for the subsequent analysis, where we introduce a mapping between the GUP deformation parameter and the fundamental momentum scale characterizing the Snyder model. Establishing this correspondence facilitates the comparison between the two approaches.

\subsection{Corrections to Friedmann equations due to area quantization}

The seminal work of Jacobson \cite{jac} revealed a deep connection between thermodynamics and gravity. In particular, it was shown that the Friedmann equations of cosmology can be derived from the first law of thermodynamics, provided that the horizon entropy satisfies the holographic principle, 
$ S= \dfrac{A}{4G}$,
where $A$ is the horizon surface area and $G$ is Newton's gravitational constant \cite{Pad1,Pad2,Frolov,CaiKim,AkbarCai,CaiCao}. 

Building on this insight, it has been realized that modifications of the entropy-area relation, such as those arising from the deformed Heisenberg algebra of the Snyder \cite{rz} or GUP \cite{aa,cos1,cos3} models, naturally lead to corrections in the Friedmann equations (see also \cite{Sheykhi:2018dpn, Saridakis:2020lrg, Luciano:2023zrx} for applications in extended entropic models of different origins). In both frameworks, the main feature is the existence of a fundamental length scale, which modifies the standard entropy-area law and, consequently, the cosmological dynamics.
For instance, in the context of the widely studied quadratic GUP $[x_i, p_j] = i\delta_{ij}(1 + \beta p^2/m_p^2)$ \cite{Kempf:1994su,GUP2,GUP3,GUP4,GUP5,GUP6,GUP7,GUP8,GUP9,Scardigli}, the modification to the Hubble parameter has been obtained in \cite{cos3} as\footnote{The difference in the numerical prefactor of the correction term between Eq.~\eqref{hmod} and Ref.~\cite{cos3} stems from the distinct definitions of the deformation parameter 
$\beta$ used in the two works.}
\begin{equation}\label{hmod}
H_{QA} = H_{GR}\left( 1 + \frac{\pi}{12} \frac{\beta}{m_P^4} \rho_0 \right),
\end{equation}
where $H_{GR}=\sqrt{\frac{8\pi G}{3}\rho_0}$ denotes the Hubble expansion rate in standard cosmology based on General Relativity, $m_p=1/\sqrt{G}$ is the Planck mass and $\beta$ is a dimensionless deformation parameter associated with the GUP. The subscript ``QA'' refers to the quantization of the area induced by effects of GUP.

In this approach, the total energy density $\rho_0$ of the perfect fluid  is assumed to retain its standard form, i.e., 
\begin{equation}
\label{denstot}
\rho_0=\rho_{BE,0}+\rho_{FD,0} = \frac{\pi^2}{30}\, T^4
\left(
\sum_{\text{bos.}} g_b
+ \frac{7}{8} \sum_{\text{ferm.}} g_f
\right),
\end{equation}
while the effects of the GUP are entirely captured by a deformation of the functional dependence of the Hubble parameter on $\rho_0$. Therefore, as a function of temperature, Eq. \eqref{hmod} becomes
\begin{equation}
\label{Hgupbis}
    H_{QA}(T)=H_{GR}(T)\left[1+\frac{\pi^3}{360}\frac{\beta}{m_p^4}\left(\sum_{\text{bos.}} g_b
+ \frac{7}{8} \sum_{\text{ferm.}} g_f
\right)T^4\right], 
\end{equation}
where, in both Eq. \eqref{denstot} and Eq. \eqref{Hgupbis}, we have separated the contribution of the bosonic and fermionic degrees of freedom. 

To establish a mapping between the quadratic GUP model and the Snyder framework, a commonly used alternative parameterization of the GUP parameter in the Snyder notation is $\beta = \frac{m_P^2}{\mathcal{P}^2}$ (see Eq. \eqref{formal} in Appendix \ref{AppA}). 
Clearly, in the limit $\beta \to 0$ (equivalently, $\mathcal{P} \to \infty$), the standard Heisenberg uncertainty principle is recovered, 
reproducing the usual Friedmann evolution.



\subsection{Corrections from momentum-space curvature}

We now turn to the calculation of the modified Hubble rate in the context of the Snyder model. Within this framework, the approach is fundamentally different, since the corrections arise from a modification of the energy density of the perfect fluid, rather than from a deformation of the Friedmann equations themselves. The resulting expression for the Hubble parameter will then be compared with the one obtained in the GUP framework in Eq.~\eqref{hmod}, by making use of the mapping~\eqref{formal} between the GUP parameter \( \beta \) and the Snyder momentum scale \( \mathcal{P} \).

Using Eqs.~\eqref{densB} and~\eqref{densF}, after straightforward algebra, one finds that
\begin{eqnarray}
\nonumber
    H_{MC}(T)&=&\sqrt{\frac{8\pi\rho(T)}{3m_p^2}}\\[2mm]
    &=&H_{GR}(T)\left[1-\frac{20\pi^2}{63}\frac{\left(\sum_{\text{bos.}} g_b
+ \frac{31}{32} \sum_{\text{ferm.}} g_f
\right)}{\left(\sum_{\text{bos.}} g_b
+ \frac{7}{8} \sum_{\text{ferm.}} g_f
\right)}\frac{T^2}{\mathcal{P}^2}+\mathcal{O}\left(\frac{T^4}{\mathcal{P}^4}\right)
    \right],
    \label{Hsn}
\end{eqnarray}
where the subscript ``MC'' now refers to the curvature of momentum space that emerges in the Snyder model.

From the comparison with Eq. 
\eqref{Hgupbis}, it follows that the two modified Hubble rates give rise to qualitatively different corrections with respect to the standard expression $H_0$. In the Snyder framework, the leading deviation is negative, as a direct consequence of the reduction of the energy density with respect to the unperturbed case,
and scales as $T^2/\mathcal{P}^2$. By contrast, in the GUP-modified scenario the correction is positive and arises at order $T^4/m_P^4$.

\subsection{Big Bang Nucleosynthesis}

Big Bang Nucleosynthesis (BBN) describes the sequence of nuclear reactions that led to the formation of the primordial light elements, including hydrogen (H), its isotope deuterium (D), the helium isotopes $^3$He and $^4$He and lithium $^7$Li~\cite{Kolbe,Alpher:1948ve,Bern}. This process is believed to have occurred in the early Universe, shortly after the Big Bang, once the cosmic plasma had cooled sufficiently to allow the formation of stable protons and neutrons.
Because BBN directly determines the primordial chemical composition of the Universe, the predicted abundances of these light elements are subject to stringent observational constraints. As a consequence, BBN provides one of the most powerful and sensitive probes for testing and constraining non-standard cosmological scenarios.

In particular, in the following analysis we shall constrain the parameters characterizing cosmology based on the Snyder model by requiring consistency between its cosmological predictions and key observational bounds from BBN. Specifically, we impose agreement with the existing upper limits on deviations of the freeze-out temperature from its standard value, ensuring that the synthesis of light elements remains compatible with observations.

Toward this end, we assume that the cosmic evolution follows the modified expansion rate \eqref{Hsn} predicted by the Snyder model. Furthermore, we consider that the energy density is dominated by ultrarelativistic particles, namely photons, electron-positron pairs and the three neutrino species. Photons, as massless bosons, contribute two degrees of freedom corresponding to their two polarization states, while the fermionic sector includes electrons and positrons, each with two spin states, and the three neutrino species, each contributing two degrees of freedom in the relativistic limit. Altogether, the fermions contribute ten degrees of freedom. 

Using these values, the combinations of bosonic and fermionic contributions that determine the leading corrections to the Hubble expansion in the Snyder framework can be explicitly evaluated as
\begin{equation}
    \sum_{\text{bos.}} g_b
+ \frac{7}{8} \sum_{\text{ferm.}} g_f\approx10.75\,, \qquad\,\, \sum_{\text{bos.}} g_b
+ \frac{31}{32} \sum_{\text{ferm.}} g_f\approx 11.69\,.
\end{equation}

According to the standard BBN scenario, free neutrons and protons became abundant shortly after the QCD confinement transition, at cosmic times
$t \sim 10^{-5}\!-\!10^{-4}\,\mathrm{s}$, once the temperature of the Universe had dropped sufficiently.
During the subsequent evolution - from the first fractions of a second up to several minutes - the abundances of the lightest atomic nuclei were established.
As discussed before, throughout this epoch the energy and number densities of the Universe were dominated by ultrarelativistic leptons (electrons, positrons, and neutrinos) and photons.
Owing to their rapid interactions, these species remained in thermal equilibrium, so that $T_\nu = T_e = T_\gamma \equiv T_{\rm eq}$,
where $T_{\rm eq}$ denotes the common plasma temperature, typically
$T \simeq 1\!-\!2~\mathrm{MeV}$~\cite{Bern}.
As the Universe cooled further, nuclear reactions became efficient,
leading to the formation of light nuclei, most notably primordial $^4\mathrm{He}$.

In contrast, the relatively sparse populations of neutrons and protons were kept in chemical equilibrium
through weak interactions with the leptonic plasma. 
\label{b}
As the temperature decreases, however, these interaction rates drop steeply, and the interconversion time scale eventually becomes comparable to the expansion time scale of the Universe.
In this regime, departures from chemical equilibrium start to develop, signalling the onset of the freeze-out of the weak interactions.
In order to obtain analytic estimates of these rates in the freeze-out regime, we follow~\cite{Bern,Ghoshal:2021ief,Capozziello:2017bxm} and express the total rate as
\begin{equation}
\Lambda(T) = 4 A T^3 \left( 4! T^2 + 2 \times 3! Q T + 2! Q^2 \right),
\label{Lambda}
\end{equation}
where $A=1.02\times 10^{-11}\,\mathrm{GeV}^{-4}$ and $Q=m_n-m_p$ is the mass difference of neutron and proton.

The primordial $^4\mathrm{He}$ mass fraction of the total baryonic mass
can be estimated as~\cite{Kolbe,cos3,Ghoshal:2021ief}
\begin{equation}
Y_p \equiv \frac{2\,x(t_f)}{1 + x(t_f)}\, \gamma,
\end{equation}
where
$x(t_f) =  e^{-Q/T(t_f)}$ is the neutron-to-proton equilibrium ratio at the time of weak interaction freeze-out $t_f\approx 1\,\mathrm{s}$ and $\gamma = e^{-(t_n - t_f)/\tau} \simeq 1$ accounts for the fraction of neutrons that survive until the onset of nucleosynthesis at $t_n\approx 200\,\mathrm{s}$, with $\tau \simeq 877~\mathrm{s}$ being the neutron mean lifetime.

Deviations of the $^4\mathrm{He}$ mass fraction due to a variation in
the freeze-out temperature $T_f$ can be expressed as~\cite{cos3,Capozziello:2017bxm}
\begin{equation}
\label{dyp}
\delta Y_p = Y_p \left[ \left( 1 - \frac{Y_p}{2\gamma} \right)
\log \left( \frac{2\gamma}{Y_p} - 1 \right) - \frac{2 t_f}{\tau} \right]
\frac{\delta T_f}{T_f},
\end{equation}
where $\delta T (t_n)$ has been set to zero because the nucleosynthesis
temperature $T_n$ is determined by the deuterium binding energy~\cite{Lambiase:2012fv,Capozziello:2017bxm}.
This expression quantifies the sensitivity of the helium mass fraction
to small changes in the freeze-out temperature, taking into account
neutron decay between freeze-out and nucleosynthesis.

We use the following data as a baseline to constrain the Snyder parameter \cite{Part}:
\begin{equation}
    Y_p= 0.245 \,,\qquad |\delta Y_p|\lesssim10^{-4}\,.
\end{equation}
Inserting these numerical values into Eq. \eqref{dyp}, one infers the upper bound
\begin{equation}
\label{expb}
    \left|\frac{\delta T_f}{T_f}\right|\lesssim 10^{-4}\,.
\end{equation}

The freeze-out temperature is defined implicitly as the temperature at which the weak
interaction rate governing neutron-proton interconversion equals the Hubble expansion. In the standard cosmological model, this temperature is approximately $T_{f,0}\approx 0.6\,\mathrm{MeV}$, which
follows from the standard computation with $H_{GR}(T_{f,0})=\Lambda(T_{f,0})$.
On the other hand, let us apply the same condition to the modified Hubble rate in Eq.~\eqref{Hsn}. We need to solve the equation
\begin{equation}\label{fineq}
    \alpha\left(1-\xi \frac{T^2}{\mathcal P^2}\right)=8 A T \left( 12 T^2 + 6 Q T +  Q^2 \right),
\end{equation}
where we have defined \begin{equation}
\alpha\equiv\dfrac{2\pi^{3/2} \sqrt{\frac{\sum_{\text{bos.}} g_b
+ \frac{7}{8} \sum_{\text{ferm.}} g_f}{5}}}{3m_p}\approx\frac{5.44}{m_p}\,,\qquad \xi\equiv\frac{20\pi^2}{63}\frac{\left(\sum_{\text{bos.}} g_b
+ \frac{31}{32} \sum_{\text{ferm.}} g_f
\right)}{\left(\sum_{\text{bos.}} g_b
+ \frac{7}{8} \sum_{\text{ferm.}} g_f
\right)}\approx 3.41\,.
\end{equation}

For consistency with the order of the expansion retained so far, we adopt the ansatz $T_f=T_{f,0}\left(1+c\dfrac{T_{f,0}^2}{\mathcal{P}^2}\right)$, with $c$ a constant to be determined. By substituting into Eq. \eqref{fineq}, we obtain 
\begin{equation}
    c=- \frac{\left(Q^2 + 6 Q T_{f,0} + 12 T_{f,0}^2\right) \, \xi}{(Q + 6 T_{f,0})^2}\,,
\end{equation}
where we have used the zero-th order equality $\alpha=8 AT_{f,0}\left(12 T_{f,0}^2+6 Q T_{f,0}+Q^2\right)$. In turn, the temperature fluctuation takes the form
\begin{equation}
\label{dtovT}
    \frac{\delta T_f}{T_f}
    =\frac{T_{f,0}^2}{\mathcal{P}^2} \frac{\left(Q^2 + 6 Q T_{f,0} + 12 T_{f,0}^2\right)\xi}{\left(Q + 6 T_{f,0}\right)^2}+\mathcal{O}\left(\frac{T^4_{f,0}}{\mathcal{P}^4}\right)\,.
\end{equation}
It is worth noting that the above relation has been derived using the standard form \eqref{Lambda} of the weak interaction rate. Indeed, although the Snyder deformation modifies the phase-space measure, its impact on $\Lambda(T)$ in the freeze-out regime $T\ll \mathcal{P}$ is negligible. This can be understood by noting that, while the Hubble rate depends on the total energy density through an integral over the full momentum distribution, the freeze-out interaction rates - dominated by energies much lower than $\mathcal{P}$ - receive only subleading corrections. This justifies retaining Snyder modifications in $H(T)$ while neglecting them in the interaction rates. Nevertheless, even if one were to include Snyder-induced effects to $\Lambda(T)$ at order $\mathcal{O}(T^2/\mathcal{P}^2)$, these would merely modify the overall numerical prefactor, without altering the functional dependence of the result \eqref{dtovT} on $1/\mathcal{P}^2$. Consequently, the parametrization $\eqref{Lambda}$ remains a good approximation for our purposes.

\begin{figure}[t]
\centering\includegraphics[width=0.6\textwidth]{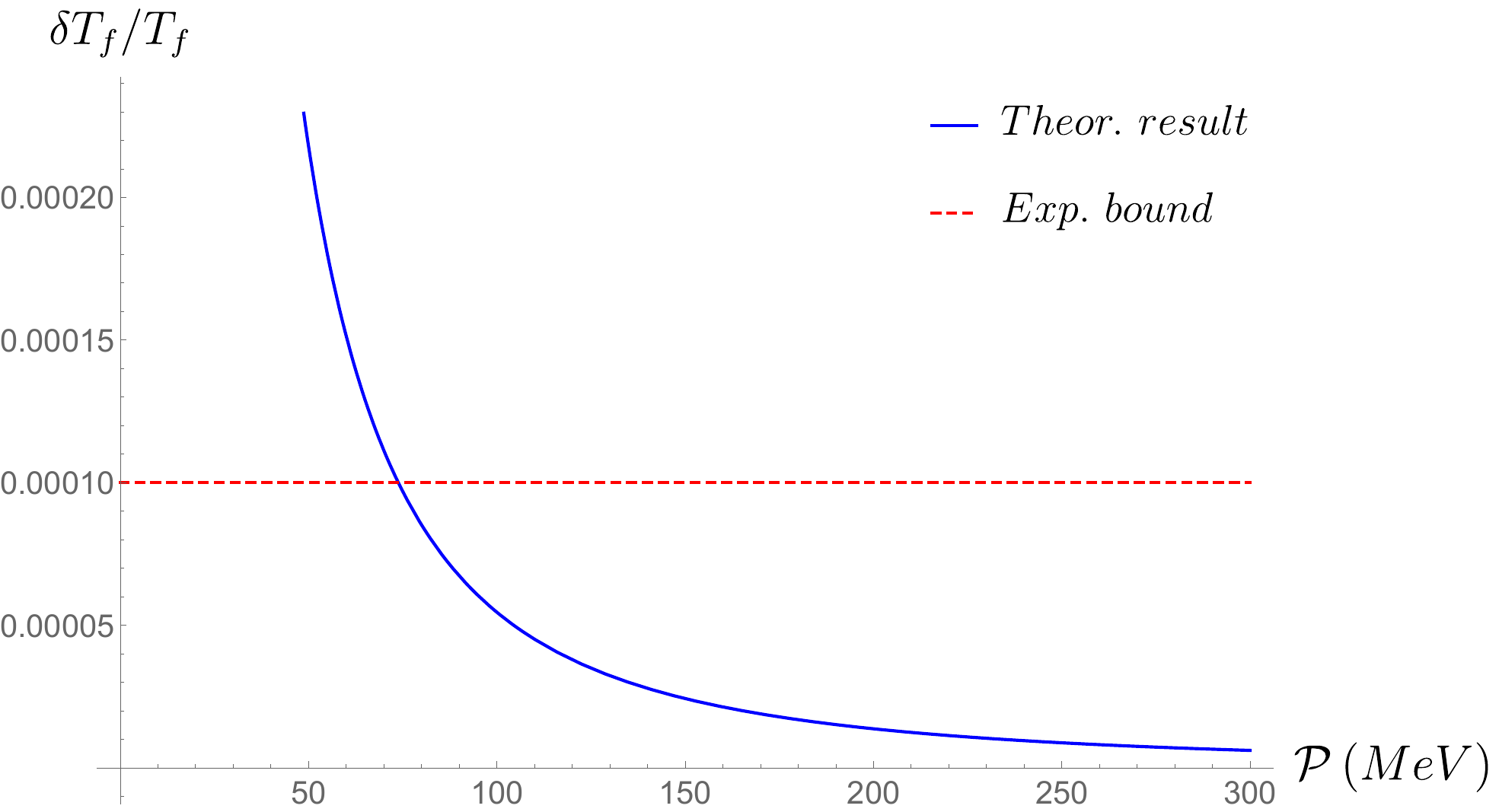}
\caption{Relative variation $\delta T_f/T_f$ as a function of $\mathcal P$. 
The solid blue line shows the theoretical prediction from Eq.~\eqref{dtovT}, 
while the red dashed line corresponds to the experimental bound from Eq.~\eqref{expb}. 
We set $T_{f,0}=0.6\,\mathrm{MeV}$ and $Q=1.293\,\mathrm{MeV}$.}
\label{Fig1}%
\end{figure}

Lastly, using the constraint \eqref{expb}, we infer the lower bound $\mathcal{P}\gtrsim 10^2\,\mathrm{MeV}$ (see also Fig. \ref{Fig1}). 
We note that this bound is not particularly restrictive. This outcome is, in fact, expected, since the energy scale relevant for BBN, of the order of a few tenths of MeV, lies well below the scales at which Snyder-induced corrections are typically anticipated to become significant.
Nevertheless, in order to facilitate comparison with existing results in the literature and to extend the scope of the present analysis, this bound can be exploited to  obtain meaningful insights within the GUP framework. This is achieved by employing the mapping~\eqref{formal} between the GUP parameter $\beta$ and the Snyder scale $\mathcal{P}$, which leads to
\begin{equation}
\beta=m_p^2/\mathcal{P}^2\lesssim 10^{40}\,.
\end{equation}
Remarkably, this bound represents a substantial improvement - by roughly 37 orders of magnitude - over the constraint previously reported in Ref.~\cite{cos3}, derived within the same BBN setting. More broadly, it stands among the most stringent constraints on the GUP parameter currently available from cosmological and astrophysical considerations\footnote{As observed in \cite{cos2,Bosso:2023aht, cos6}, 
a number of studies have explored possible bounds on the GUP parameter within low-energy quantum frameworks. Although some of these analyses yield very tight limits, direct comparisons across different approaches should be treated with care, since the GUP can manifest itself in distinct physical systems through mechanisms that admit different theoretical interpretations~\cite{Bosso:2023aht}. For this reason, in the present work we restrict our comparative discussion to constraints obtained in cosmological and astrophysical settings that are closely aligned with the scenario investigated here.} \cite{cos2,Bosso:2023aht}.
For comparison, we recall that the analysis of black hole shadows yields an upper bound 
$\beta \lesssim 10^{90}$ \cite{Neves}, while a tighter constraint, 
$\beta \lesssim 10^{69}$, is inferred from Solar System measurements of the perihelion 
precession~\cite{Scardigli:2014qka}. These limits are further improved to 
$\beta \lesssim 10^{59}$ when full cosmological datasets are taken into 
account~\cite{cos2}. Even stronger bounds have been reported in scenarios involving 
violations of the Equivalence Principle~\cite{Eqp}. On the other hand, 
focusing on gravitational-wave  probes, studies of modified graviton dispersion 
relations lead to constraints of the order $\beta \lesssim 10^{60}$~\cite{Feng}, or to 
comparable bounds at the level of $\beta \lesssim 10^{39}$ in alternative GW-based 
analyses~\cite{cos6}. 

Therefore, our results underscore the potential role of high-precision cosmology as a powerful probe of non-commutative frameworks such as the Snyder model and the GUP, highlighting the importance of cosmological and astrophysical observations - particularly those involving early-Universe processes - in advancing our understanding of Planck-scale physics.

\section{Conclusion and outlook}
\label{Conc}

In this work, we have presented a comprehensive analysis of statistical mechanics on Euclidean Snyder space, emphasizing the geometric origin of noncommutativity as curvature in momentum space. By formulating all physical quantities in terms of momentum-space invariants, we have constructed a realization-independent framework that provides a robust and physically transparent extension of standard statistical mechanics.
We have explicitly derived the modified partition functions and thermodynamic quantities for Maxwell-Boltzmann, Bose-Einstein and Fermi-Dirac systems, covering both non-relativistic and ultrarelativistic regimes. A central result of our analysis is the universal suppression of entropy, internal energy and energy density induced by the compactness of Snyder momentum space. This suppression can be interpreted as a direct manifestation of the effective ultraviolet cutoff encoded in the Snyder geometry, which progressively limits the number of accessible microstates at high energies.

In the ultrarelativistic regime, relevant for the early Universe, we have obtained closed analytic expressions for the leading corrections to the energy density of relativistic matter. These corrections scale as $T^{2}/\mathcal P^{2}$ and lead to a reduced Hubble expansion rate compared to standard cosmology. This behavior stands in sharp qualitative contrast with quadratic GUP-based cosmological models, based on the quantization of area, where the leading corrections enhance the expansion rate and scale as $T^{4}/m_{p}^{4}$. Our results thus highlight that different quantum-gravity-inspired deformations, even when characterized by a minimal length, can give rise to distinct and observationally distinguishable cosmological signatures.

We have exploited the sensitivity of Big Bang Nucleosynthesis (BBN) to early-Universe dynamics to constrain the Snyder momentum scale. Requiring consistency with observational bounds on variations of the freeze-out temperature, we have derived a lower bound on $\mathcal{P}$. As discussed above, this constraint is not particularly stringent, reflecting the fact that BBN probes energy scales that are well below those at which Snyder-induced corrections are expected to play a significant role. Nevertheless, it should be emphasized that moving to higher energy regimes - such as those relevant for the electroweak phase transition, reheating dynamics after inflation, or high-energy particle production in the early Universe - one may reasonably expect this bound to be substantially strengthened.

Furthermore, by means of a phenomenological mapping between the Snyder framework and the Generalized Uncertainty Principle, we have translated the bound on $\mathcal{P}$ into a constraint on the quadratic GUP deformation parameter. Remarkably, the resulting limit improves upon existing BBN-based constraints by many orders of magnitude, placing it among the strongest bounds currently obtained from cosmological and astrophysical considerations.

Looking ahead, several aspects deserve further investigation. Extensions of the present framework to interacting systems and to quantum field theories on Snyder space would allow a more complete assessment of noncommutative effects in thermal and cosmological settings. Moreover, exploring the implications of momentum-space curvature for phase transitions, reheating and relic abundances could further sharpen the phenomenological impact of the Snyder model. Finally, confronting Snyder-induced corrections with forthcoming high-precision cosmological data may offer new opportunities to probe the geometric structure of momentum space and the imprint of Planck-scale physics on the early Universe. 
Work in these directions is presently under active investigation.

\appendix

\section{On the Snyder algebra and its implications}
\label{AppA}
In this appendix we briefly review the algebraic construction of Euclidean Snyder space. 
Position operators are defined in terms of the generators of the Lie algebra $so(4)$, leading to noncommuting spatial coordinates while preserving rotational symmetry. 
Specifically, we have
 \begin{equation}
     x_i=\frac{1}{\mathcal P} J_{i4}\,,
 \end{equation}
while the other three generators of the $so(4)$ algebra are the angular momentum operators $J_{ij}$ ($i<j)$.
This reproduces our algebra: 
 \begin{equation}
     [x_i,x_j]=i\frac{J_{ij}}{\mathcal P^2} 
 \end{equation}
 and
 \begin{equation}
   [x_i,\pi_j]=\frac{1}{\mathcal P}[J_{i4},\pi_j]=\frac{i}{\mathcal P}\delta_{ij}\pi_4={{i}}\delta_{ij} \sqrt{1-\frac{\pi^2}{\mathcal P^2}}\,,  
 \end{equation}
where, in the last relation, we have used Eq. \eqref{3surf}.
Since we are using the representation where \ \(g=1\), the physical momenta coincide with the embedding coordinates, so that \(p_i = \pi_i\) and
\begin{equation}
   [x_i,p_j]=i\delta_{ij} \sqrt{1-\frac{p^2}{\mathcal P^2}}\,,  
   \label{alg3}
 \end{equation}
 which is the algebra given in Eq. \eqref{alg2}.

The properties of the $so(4)$ algebra have been studied in great detail in Ref.~\cite{ref8}. 
Here we briefly summarize those features that are most relevant for our purposes. 
First, the Cartesian coordinate operators $x_i$ possess a discrete and evenly spaced spectrum of eigenvalues. In this sense, Snyder configurational space can be interpreted as a cubical lattice with lattice spacing 
$\mathcal{P}^{-1}$.
This lattice spacing is commonly identified with a fundamental length scale.\footnote{This effective lattice structure should be distinguished from the lattices employed in conventional lattice field theories \cite{Wilson} or in geometrically discretized space-time models, such as the crystal-like universe considered in \cite{Jizba:2009qf}. In lattice field theory, the lattice is introduced as a regularization device, providing a non-perturbative definition of quantum field theories, while Lorentz invariance is explicitly broken and recovered only in the continuum limit. In the Snyder framework, by contrast, spatial discreteness arises at the level of coordinate operators while preserving an undeformed action of the Lorentz group. Crystal-like models instead implement discreteness geometrically, leading to an explicit spatial lattice structure and modified kinematical properties.}
At the same time, the eigenvalues of the squared distance operator, $r^2 = \sum_i x_i^2$, are also discrete and evenly 
spaced, leading to a quantization of area, as also observed in Ref.~\cite{rz}. A similar outcome has been discussed in Ref.~\cite{ref7}.

Taken together, these two properties allow us to formally relate the Snyder framework to the quadratic GUP model introduced in \cite{Kempf:1994su}. 
In particular, the comparison can be made in terms of their characteristic length/area scales. 
In the GUP model, the presence of a minimal length $\Delta x_{\rm min} = \sqrt{\beta}\,\ell_p = \frac{\sqrt{\beta}}{m_P}$,
where $\ell_p$ is the Planck length and $\beta$ the GUP deformation parameter, leads naturally to an identification of the Snyder and GUP parameters via
\begin{equation}
\label{formal}
\frac{1}{\mathcal P}=\frac{\sqrt{\beta}}{m_P} \,\,\Longrightarrow\,\,\beta = \frac{m_P^2}{\mathcal{P}^2}\,.
\end{equation}
We emphasize that this correspondence is intended as a phenomenological mapping: it allows one to relate the characteristic scales of the two models for comparative purposes, but does not imply a strict equivalence between their underlying algebraic structures.

\begin{acknowledgments}
The research of GGL is supported by the postdoctoral fellowship
program of the University of Lleida. GGL gratefully acknowledges
the contribution of the LISA Cosmology Working Group (CosWG), as
well as support from the COST Actions CA21136 - \textit{Addressing
observational tensions in cosmology with systematics and
fundamental physics (CosmoVerse)} - CA23130, \textit{Bridging high
and low energies in search of quantum gravity (BridgeQG)} and
CA21106 - \textit{COSMIC WISPers in the Dark Universe: Theory,
astrophysics and experiments (CosmicWISPers)}. 
\end{acknowledgments}


\begin{thebibliography}{99}
\bibitem{ref1} C. A. Mead, Phys. Rev. B135, 849 (1964).

\bibitem{ref2} S. Doplicher, K. Fredenhagen, J. E. Roberts, Commun. Math. Phys. 172, 187 (1995).

\bibitem{ref3} E. Witten, Nucl. Phys. B \textbf{268}, 253 (1986). 

\bibitem{ref4} J. Madore et al., Eur. Phys. J. C \textbf{16}, 161 (2000).

\bibitem{ref5} N. Seiberg, E. Witten, JHEP \textbf{09}, 032 (1999).  

\bibitem{Rovelli:1994ge}
C.~Rovelli and L.~Smolin,
Nucl. Phys. B \textbf{442}, 593-622 (1995).

\bibitem{Maldacena:1997re}
J.~M.~Maldacena,
Adv. Theor. Math. Phys. \textbf{2}, 231-252 (1998).

\bibitem{Kempf:1994su}
A.~Kempf, G.~Mangano and R.~B.~Mann,
Phys. Rev. D \textbf{52}, 1108-1118 (1995).

\bibitem{Addazi:2021xuf}
A.~Addazi, J.~Alvarez-Muniz, R.~Alves Batista, G.~Amelino-Camelia, V.~Antonelli, M.~Arzano, M.~Asorey, J.~L.~Atteia, S.~Bahamonde and F.~Bajardi, \textit{et al.}
Prog. Part. Nucl. Phys. \textbf{125}, 103948 (2022).

\bibitem{Hossenfelder:2012jw}
S.~Hossenfelder,
Living Rev. Rel. \textbf{16}, 2 (2013).

\bibitem{Bosso:2023aht}
P.~Bosso, G.~G.~Luciano, L.~Petruzziello and F.~Wagner,
Class. Quant. Grav. \textbf{40}, 195014 (2023).

\bibitem{ref6} H.S. Snyder, Phys. Rev. \textbf{71}, 38 (1947).  

\bibitem{ref7} S. Mignemi, Phys. Rev. D \textbf{84}, 025021 (2011).  

\bibitem{ls2} L. Lu, A. Stern,  Nucl. Phys. B \textbf{860}, 186 (2012).

\bibitem{ref8} Z. Lu, A. Stern, Nucl. Phys. B \textbf{854}, 894 (2012).

\bibitem{ref9} B. Ivetić, Nucl. Phys. B \textbf{989}, 116129 (2023).

\bibitem{ref10} B. Ivetić, Phys. Rev. D \textbf{100}, 115047 (2019).

\bibitem{ref11} R.M. Mir-Kasimov, Phys. Lett. B \textbf{378}, 181 (1996).  

\bibitem{ref12} M.V. Battisti, S. Meljanac, Phys. Rev. D \textbf{82}, 024028 (2010).

\bibitem{ref13} B. Ivetić, Nucl. Phys. B \textbf{987}, 116085 (2023).  

\bibitem{ref14} Yu. A. Gol’fand, JETP \textbf{10}, 356 (1960). 

\bibitem{ref14bis} Yu. A. Gol’fand, 
JETP \textbf{16}, 184 (1963). 

\bibitem{ref14ter} Yu. A. Gol’fand,
JETP \textbf{17}, 842 (1963).  

\bibitem{ref15} V.G. Kadyshevsky, Nucl. Phys. B \textbf{141}, 477 (1978). 

\bibitem{ref16} V.G. Kadyshevsky, M.D. Mateev, Phys. Lett. B \textbf{106}, 139 (1981).

\bibitem{ref17} S.A. Franchino-Viñas, J.J. Relancio, Clas. Quantum Grav. \textbf{40}, 054001 (2023).  

\bibitem{ref18} B. Ivetić, Nucl. Phys. B \textbf{1007}, 116692 (2024). 

\bibitem{ref19} K. Nozari, V. Hosseinzadeh, M.A. Gorji, Phys. Lett. B \textbf{750}, 218 (2015).  

\bibitem{ref20} V. Hosseinzadeh, M.A. Gorji, K. Nozari, B. Vakili, Phys. Rev. D \textbf{92}, 025008 (2015). 

\bibitem{iran3} M.A. Gorji, V. Hosseinzadeh, K. Nozari, B. Vakili, Phys. Rev. D \textbf{93}, 064029 (2016). 

\bibitem{iran4} M.A. Gorji, V. Hosseinzadeh, K. Nozari, B. Vakili, J. Stat. Mech 073107 (2016).


\bibitem{Pachol2023}
A. Pachol, A. Wojnar, Class. Quant. Grav. \textbf{40}, 195021 (2023).

\bibitem{Pachwoj}
A. Pachol, A. Wojnar, 
 	Eur. Phys. J. C \textbf{83}, 1097 (2023).


\bibitem{Ter1}
A.~Kobakhidze, C.~Lagger and A.~Manning,
Phys. Rev. D \textbf{94}, no.6, 064033 (2016).

\bibitem{Ter2}
G.~Lambiase, M.~Sakellariadou and A.~Stabile,
JCAP \textbf{12}, 020 (2013).

\bibitem{Ter3}
P.~K.~Joby, P.~Chingangbam and S.~Das,
Phys. Rev. D \textbf{91}, no.8, 083503 (2015).

\bibitem{cos1} S. Das, M. Fridman, G. Lambiase, E.C. Vagenas, Phys. Lett. B \textbf{824}, 136841 (2022).

\bibitem{cos2} S. Giardino, V. Salzano, Eur. Phys. J. C \textbf{81}, 110 (2021).

\bibitem{cos3} G.G. Luciano, Eur. Phys. J. C \textbf{81}, 1086 (2021).

\bibitem{cos4}
K.~Zeynali, F.~Darabi and H.~Motavalli,
JCAP \textbf{12}, 033 (2012).

\bibitem{cos5}
A.~Paliathanasis,
JCAP \textbf{09}, 067 (2025).

\bibitem{cos6}
G.~G.~Luciano and Y.~Sekhmani,
Phys. Lett. B \textbf{862}, 139315 (2025).

\bibitem{cos7}
S.~Kouwn,
Phys. Dark Univ. \textbf{21}, 76-81 (2018).

\bibitem{Jizba:2009qf}
P.~Jizba, H.~Kleinert and F.~Scardigli,
Phys. Rev. D \textbf{81}, 084030 (2010).

\bibitem{Maggiore1993}
M. Maggiore, Phys. Lett. B \textbf{319}, 83 (1993).


\bibitem{Carmona2019}
J.M. Carmona, J.L. Cortes, J.J. Relancio, Phys. Rev. D \textbf{100}, 104031 (2019).











%
%
%
%
%
%

\bibitem{Kolbe}
E.~W.~Kolb and M.~S.~Turner,
Front. Phys. \textbf{69}, 1-547 (1990).

\bibitem{Landau}
L.~D.~Landau and E.~M.~Lifshitz,
Butterworth-Heinemann, 1980.

\bibitem{jac} T. Jacobson, Phys. Rev. Lett. \textbf{75}, 1260 (1995).

\bibitem{Pad1} T. Padmanabhan, Phys. Rept. \textbf{406}, 49 (2005).

\bibitem{Pad2} T. Padmanabhan, Rept. Prog. Phys. \textbf{73}, 046901 (2010).

\bibitem{Frolov} A. V. Frolov and L. Kofman, JCAP \textbf{05}, 009 (2003).

\bibitem{CaiKim} R.-G. Cai and S. P. Kim, JHEP \textbf{02}, 050 (2005).

\bibitem{AkbarCai} M. Akbar and R.-G. Cai, Phys. Rev. D \textbf{75}, 084003 (2007).

\bibitem{CaiCao} R.-G. Cai and L.-M. Cao, Phys. Rev. D \textbf{75}, 064008 (2007).


\bibitem{rz}
J.M. Romero, A. Zamora, Phys. Lett. B \textbf{661}, 11 (2008).

\bibitem{aa} A. Awad, A.F. Ali, JHEP \textbf{06}, 093 (2014).


\bibitem{Sheykhi:2018dpn}
A.~Sheykhi,
Phys. Lett. B \textbf{785}, 118-126 (2018).

\bibitem{Saridakis:2020lrg}
E.~N.~Saridakis,
JCAP \textbf{07}, 031 (2020).

\bibitem{Luciano:2023zrx}
G.~G.~Luciano,
Phys. Lett. B \textbf{838}, 137721 (2023).


\bibitem{GUP2}
K.~Nozari and T.~Azizi,
Gen. Rel. Grav. \textbf{38}, 735-742 (2006).

\bibitem{GUP3}
S.~Das and E.~C.~Vagenas,
Phys. Rev. Lett. \textbf{101}, 221301 (2008).

\bibitem{GUP4}
P.~Pedram, K.~Nozari and S.~H.~Taheri,
JHEP \textbf{03}, 093 (2011).

\bibitem{GUP5}
F.~Scardigli,
Phys. Lett. B \textbf{452}, 39-44 (1999).

\bibitem{GUP6}
L.~Buoninfante, G.~G.~Luciano and L.~Petruzziello,
Eur. Phys. J. C \textbf{79}, no.8, 663 (2019).

\bibitem{GUP7}
S.~Hossenfelder,
Living Rev. Rel. \textbf{16}, 2 (2013).

\bibitem{GUP8}
R.~J.~Adler, P.~Chen and D.~I.~Santiago,
Gen. Rel. Grav. \textbf{33}, 2101-2108 (2001).

\bibitem{GUP9}
S.~Capozziello, G.~Lambiase and G.~Scarpetta,
Int. J. Theor. Phys. \textbf{39}, 15-22 (2000).

\bibitem{Scardigli}
M.~Blasone, G.~Lambiase, G.~G.~Luciano, L.~Petruzziello and F.~Scardigli,
Int. J. Mod. Phys. D \textbf{29}, no.02, 2050011 (2020).

\bibitem{Alpher:1948ve}
R.~A.~Alpher, H.~Bethe and G.~Gamow,
Phys. Rev. \textbf{73}, 803-804 (1948).

\bibitem{Bern}
J.~Bernstein, L.~S.~Brown and G.~Feinberg,
Rev. Mod. Phys. \textbf{61}, 25 (1989).

\bibitem{Ghoshal:2021ief}
A.~Ghoshal and G.~Lambiase,
[arXiv:2104.11296 [astro-ph.CO]].


\bibitem{Capozziello:2017bxm}
S.~Capozziello, G.~Lambiase and E.~N.~Saridakis,
Eur. Phys. J. C \textbf{77}, no.9, 576 (2017).

\bibitem{Lambiase:2012fv}
G.~Lambiase,
JCAP \textbf{10}, 028 (2012).

\bibitem{Part}
S.~Navas \textit{et al.} [Particle Data Group],
Phys. Rev. D \textbf{110}, no.3, 030001 (2024).

\bibitem{Neves}
J.~C.~S.~Neves,
Eur. Phys. J. C \textbf{80}, no.8, 717 (2020)

\bibitem{Scardigli:2014qka}
F.~Scardigli and R.~Casadio,
Eur. Phys. J. C \textbf{75}, no.9, 425 (2015).

\bibitem{Eqp}
S.~Ghosh,
Class. Quant. Grav. \textbf{31}, 025025 (2014).

\bibitem{Feng}
Z.~W.~Feng, S.~Z.~Yang, H.~L.~Li and X.~T.~Zu,
Phys. Lett. B \textbf{768}, 81-85 (2017).

\bibitem{Wilson}
K.~G.~Wilson,
Phys. Rev. D \textbf{10}, 2445-2459 (1974).










\end{thebibliography}
\end{document}